\documentclass[12pt]{iopart}

\usepackage{graphicx}
\begin{document}

\title[]{The GUP effect on Hawking Radiation of the 2+1 dimensional Black Hole}

\author{Ganim Gecim and Yusuf Sucu}

\address{Department of Physics, Faculty of Science,
Akdeniz University, \\ 07058 Antalya, Turkey}
\ead{gecimganim@gmail.com and ysucu@akdeniz.edu.tr} \vspace{10pt}

\begin{abstract}
We investigate the Generalized Uncertainty Principle (GUP) effect on
the Hawking radiation of the 2+1 dimensional Martinez-Zanelli black
hole by using the Hamilton-Jacobi method. In this connection, we
discuss the tunnelling probabilities and Hawking temperature of the
spin-1/2 and spin-0 particles for the black hole. Therefore, we use
the modified Klein-Gordon and Dirac equations based on the GUP.
Then, we observe that the Hawking temperature of the scalar and
Dirac particles depend on not only the black hole properties, but
also the properties of the tunnelling particle, such as angular
momentum, energy and mass. And, in this situation, we see that the
tunnelling probability and the Hawking radiation of the Dirac
particle is different from that of the scalar particle.
\end{abstract}


\maketitle

\section{Introduction}\label{intro}

The discovery of the black hole radiation, known as Hawking
radiation in the literature, is one of the milestones to construct a
consistent connection between the relativity theory, the statistical
mechanics and the quantum mechanics. The nature of a black hole has
been started to be investigated in the framework of the
thermodynamical and the quantum mechanical concepts since 1970
\cite{1,2,3,4,5,6}. Hawking investigated the thermodynamical
properties of a black hole in the frame of quantum field theory
based on the Heisenberg uncertainty principle on a curved spacetime.
Since then, the Hawking radiation has been investigated as a quantum
tunnelling effect of the relativistic particles from a black hole
\cite{9,9a,9b,10,11,11b,12,13}. Also, the Hawking radiation as a
tunnelling process of the particles from various black holes has
been studied, extensively, in the literature in both 3+1 and 2+1
dimensional \cite{12,13,14,15,16,17,18,19,19a,20}.

On the other hand, the suitable candidate quantum gravity theories,
such as string theory and loop quantum gravity theory, indicate the
presence of a minimal observable length in Planck scale
\cite{21,22,23,24}. The existence of such a minimal length leads to
the generalized Heisenberg uncertainty principle (GUP). The GUP can
be expressed as \cite{24a,24b},
\begin{eqnarray}
\Delta x\Delta p\geq \frac{h}{2}\left[ 1+\beta (\Delta
p)^{2}\right]\label{GUP1}
\end{eqnarray}
where, $\beta=\beta_{0}/ M_{p}^2$, the $M_{p}^2$ is the Planck mass
and $\beta_{0}$ is the dimensionless parameter. Then, the modified
commutation relation becomes,
\begin{eqnarray}
\left[ x_{\mu},p_{\nu}\right] =i\hbar\delta _{\mu \nu}\left[ 1+\beta
p^{2}\right]\label{GUP2},
\end{eqnarray}
where, $x_{\mu}$ and $p_{\mu}$ are the modified position and the
momentum operators, respectively, defined by
\begin{eqnarray}
x_{\mu } &=&x_{0\mu } \nonumber \\
p_{\mu } &=&p_{0\mu}(1+\beta p_{0\mu}^{2}) \label{GUP3},
\end{eqnarray}
where, the $x_{0\mu}$ and $p_{0\mu}$ are the standard position and
momentum operators, respectively, and they satisfy the usually
commutation relation $\left[ x_{0\mu},p_{0\nu}\right]
=i\hbar\delta_{\mu \nu}$. These modified relations play an important
role in physics. For example, in recent years, using the GUP, the
thermodynamics properties of the black holes were investigated via a
particle tunnelling from the black holes. To include the quantum
gravity effect, the Klein-Gordon and Dirac equations are modified by
the GUP framework \cite{NOZ1}. With these modified relativistic wave
equations, the corrected Hawking temperature of various $3+1$ and
higher dimensional black holes computed via a particle tunnelling
process \cite{24c, 24d, 24e, 24f, 24g, 24h, 24j, 24k, 24l, 24m,
24n}. In this motivation, we will investigate the Hawking radiation
of the 2+1 dimensional Martinez-Zanelli black hole by the scalar and
Dirac particles tunnelling process under the effect of the GUP. The
metric of the Martinez-Zanelli black hole is given by \cite{MZ1}
\begin{eqnarray}
ds^{2}=F(r)dt^{2}-\frac{1}{F(r)}dr^{2}-r^{2}d\theta ^{2} \label{MZ}
\end{eqnarray}
where $F(r)$,
\begin{eqnarray*}
F(r)=\frac{1}{l^{2}}\left[ r^{2}-3B^{2}-\frac{2B^{3}}{r}\right] =\frac{%
\left( r+B\right) ^{2}\left( r-2B\right) }{rl^{2}},
\end{eqnarray*}
and $l^{2}=-1/\Lambda$ is the cosmological constant and $B$ is the
mass parameter related to the black hole mass $M$, as
$B=\sqrt{Ml^{2}/3}$ \cite{MZ2}. Hence, the black hole has a
singularity at $r=0$ surrounded by horizon located at $r_{h}=2B$
under the condition $B\neq0$.

The organization of this work are as follows: In the Section 2, we
modified the Klein-Gordon equation by using the GUP. Subsequently,
from the modified Klein-Gordon equation written in the $2+1$
dimensional Martinez-Zanelli Black hole background, we calculate the
tunnelling possibility of the scalar particle by using the
semi-classical method, and then, we find the Hawking temperature. In
the Section 3, the modified Dirac equation is written in the $2+1$
dimensional Martinez-Zanelli black hole, and then, the tunnelling
probability of the Dirac particle from the black hole and its
Hawking temperature is also calculated. Finally, in conclusion, we
evaluate and summarize the results.

\section{The Modified Klein-Gordon Equation and the Scalar particle tunnelling}

To investigate the quantum gravity effect on the tunnelling process
of the scalar particles from the black hole and on its Hawking
temperature, we will discuss the modified Klein-Gordon equation
under the GUP relations. The standard Klein-Gordon equation can be
written as \cite{Greiner},
\begin{eqnarray}
p_{0\mu }p_{0}^{\mu }\phi=m_{0}^{2}\phi, \label{KG1}
\end{eqnarray}
or its explicit form is
\begin{eqnarray}
-\left(i\hbar\right) ^{2}\partial_{t}\partial^{t}\phi =\left[
\left(-i\hbar\right) ^{2}\partial _{i}\partial
^{i}-m_{0}^{2}\right]\phi, \label{KG2}
\end{eqnarray}
where $\phi$ is the wave function of the scalar particles. On the
other hand, in the context of the GUP, the modified energy relation
is given by
\begin{eqnarray}
\widetilde{E}=E\left( 1-\beta E^{2}\right) =E\left[ 1-\beta \left(
p^{2}+m_{0}^{2}\right) \right] \label{EN}
\end{eqnarray}
where $E^{2}=p^{2}+m_{0}^{2}$. Then, the square of the momentum
operator can obtained by using the Eq.(\ref{GUP3}) as follows;
\begin{eqnarray}
p^{2}=p_{\mu }p^{\mu }\simeq-\hbar^{2}\left[ \partial _{i}\partial
^{i}-2\beta \left(
\partial _{j}\partial ^{j}\right) \left( \partial _{j}\partial ^{j}\right)
\right]\label{MO}
\end{eqnarray}
where the higher order terms of the $\beta$ parameter are neglected.
Then, using the Eq.(\ref{EN}) and Eq.(\ref{MO}) in the standard
Klein-Gordon equation, the modified Klein-Gordon equation is written
as follows;
\begin{eqnarray}
-\left( i\hbar\right) ^{2}\partial _{t}\partial ^{t}\Phi =\left[
\left( -i\hbar\right) ^{2}\partial _{i}\partial
^{i}-m_{0}^{2}\right] \left[ 1-2\beta \left( -\hbar^{2}\partial
_{i}\partial ^{i}+m_{0}^{2}\right) \right] \Phi, \label{MKG1}
\end{eqnarray}
where $\Phi$ is the generalized wave function of the scalar
particles. Hence, the modified Klein-Gordon equation in the
Martinez-Zanelli black hole background is\\
\begin{eqnarray}
\frac{\hbar^{2}}{F(r)}\frac{\partial ^{2}\Phi }{\partial
t^{2}}-\hbar^{2}F(r)\frac{\partial ^{2}\Phi }{\partial
r^{2}}-\frac{\hbar^{2}}{r^{2}}\frac{\partial ^{2}\Phi }{\partial
\phi ^{2}}+2\beta F(r)\hbar^{4}\frac{\partial ^{2}}{\partial
r^{2}}\left[ F(r)\frac{\partial ^{2}\Phi }{\partial r^{2}}\right] \nonumber \\
+\frac{2\beta \hbar^{4}}{r^{2}}\frac{\partial ^{2}}{\partial \phi
^{2}}\left[ \frac{1}{r^{2}}\frac{\partial ^{2}\Phi }{\partial \phi
^{2}}\right] +m_{0}^{2}\left( 1-2\beta
m_{0}^{2}\right)\Phi =0. \label{MKG2}
\end{eqnarray}

To investigate the tunnelling radiation of the Martinez-Zanelli
black hole with the Eq.(\ref{MKG2}), we employ the wave function of
the scalar particle as,
\begin{eqnarray}
\Phi \left( t,r,\phi \right) =A e^{\frac{i}{\hbar}S\left( t,r,\phi
\right)} \label{ansatz}
\end{eqnarray}
where $A$ is a constant and $S(t,r,\phi)$ is the classical action
term for the outgoing particle trajectory. Substituting the
Eq.(\ref{ansatz}) into the Eq.(\ref{MKG2}) and neglecting the the
higher order terms of $\hbar$, we get the equation of motion of the
scalar particle as
\begin{eqnarray}
\left( \frac{\partial S}{\partial t}\right) ^{2}-F^{2}(r)\left( \frac{%
\partial S}{\partial r}\right) ^{2}-\frac{F(r)}{r^{2}}\left( \frac{\partial S%
}{\partial \phi }\right)^{2}-2\beta F^{3}(r)\left( \frac{\partial S}{%
\partial r}\right)^{4}\nonumber \\-\frac{2\beta F(r)}{r^{4}}\left( \frac{\partial S}{%
\partial \phi }\right)^{4}-m_{0}^{2}\left( 1-2\beta m_{0}^{2}\right)
F(r)=0. \label{an1}
\end{eqnarray}
Due to the commuting Killing vectors $\left( \partial _{t}\right)$
and $\left(\partial _{\phi }\right)$ we can separate the
$S\left(t,r,\phi \right)$, in terms of the variables $t$, $r$ and
$\phi$, such as $S\left( t,r,\phi \right) =-Et+j\phi +K(r)$, where
$E$ and $j$ are the energy and angular momentum of the particle,
respectively, and $K(r)=K_{0}(r)+\beta K_{1}(r)$ \cite{24k}. And,
from the Eq.(\ref{an1}), the radial integral, $K(r)$, becomes as
follows;
\begin{eqnarray*}
K_{\pm }(r) =\pm \int \frac{\sqrt{E^{2}-F(r)\left( m_{0}^{2}+\frac{j^{2}}{%
r^{2}}\right) }}{F(r)}\left[ 1+\beta \Sigma\right] dr
\end{eqnarray*}
and it is computed as,
\begin{eqnarray*}
K_{\pm }(r_{h})=\pm i\pi \frac{l^{2}E}{972B^{3}}\left[
216B^{2}+\beta \left( 324m_{0}^{2}B^{2}+16E^{2}l^{2}+81j^{2}\right)
\right]
\end{eqnarray*}
where, $\Sigma$ is
\begin{eqnarray}
\Sigma= \left(\frac{F(r)\left(m_{0}^{2}-\frac{%
j^{4}}{r^{4}}\right) }{E^{2}-F(r)\left(
m_{0}^{2}+\frac{j^{2}}{r^{2}}\right)
}-\frac{E^{2}-F(r)\left( m_{0}^{2}+\frac{j^{2}}{r^{2}}\right) }{F(r)}\right)
\end{eqnarray}
and $K_{+}(r_{h})$ is outgoing and $K_{-}(r_{h})$ is incoming
solutions of the radial part. The total imaginary part of the action
is $ImS\left(t,r,\phi \right) =ImK_{\pm }\left( r\right)
=ImK_{+}\left( r\right) -ImK_{-}\left( r\right) $ . Hence, the two
kind probabilities of the crossing from the outer horizon, from
outside to inside and from inside to outside, are given by
\cite{12,17,Volovik}
\begin{eqnarray}
P_{out}=\exp \left(-\frac{2}{h}Im K_{+}(r_{h})\right) \label{proba1}
\end{eqnarray}
and
\begin{eqnarray}
P_{in}=\exp \left(-\frac{2}{h}Im K_{-}(r_{h})\right), \label{proba2}
\end{eqnarray}
respectively. Then, the tunneling probability of the scalar particle
is written as
\begin{eqnarray*}
\Gamma =\frac{P_{out}}{P_{in}}=\exp \left\{ -\frac{\pi l^{2}E}{243\hbar B^{3}}%
\left[ 216B^{2}+\beta \left( 324m_{0}^{2}B^{2}+16E^{2}l^{2}+81j^{2}\right) %
\right] \right\}.
\end{eqnarray*}
Hence, the modified Hawking temperature is obtained from the lowest
order in the expansion of the classical action in terms of the
particle energy,
\begin{eqnarray}
\Gamma =\exp \left(-\frac{2}{\hbar}ImS\right) =\exp
\left(-\frac{E}{T_{H}^{^{\prime }}} \right)\label{proba2}
\end{eqnarray}
where $T_{H}^{^{\prime }}$ is the modified Hawking temperature of
the outer horizon, and it is given by
\begin{eqnarray*}
T_{H}^{^{\prime }}=\frac{9 \hbar B}{8\pi l^{2}}\left[ 1+\beta \frac{%
324m_{0}^{2}B^{2}+16E^{2}l^{2}+81j^{2}}{216B^{2}}\right] ^{-1}.
\end{eqnarray*}
If, at first, we expand the $T_{H}^{^{\prime }}$ in terms of the
$\beta$ powers, and, second, neglect the higher order of the $\beta$
terms, then we get the modified Hawking temperature of the
Martinez-Zanelli black hole as follows;
\begin{eqnarray}
T_{H}^{^{\prime }}= T_{H}\left[ 1-\beta \frac{%
324m_{0}^{2}B^{2}+16E^{2}l^{2}+81j^{2}}{216B^{2}}\right]
\label{KGTH}
\end{eqnarray}
where the $T_{H}=\frac{9\hbar B}{8\pi l^{2}}$ is the standard
Hawking temperature of the black hole. From the $T_{H}^{^{\prime }}$
expression, we see that the modified Hawking temperature is related
to not only the mass parameter of the black hole, but also the
angular momentum, energy and mass of the emitted scalar particle
from the black hole, and it is lower than the standard Hawking
temperature.

\section{The Modified Dirac Equation and Fermion tunnelling}

The Dirac equation in a (2+1) dimensional spacetime is given by the
following representation \cite{25},
\begin{eqnarray}
\left\{ i\overline{\sigma }^{\mu }(x)\left[ \partial _{\mu }-\Gamma
_{\mu }(x)\right] \right\} \Psi (x)=\frac{m_{0}}{\hbar }\Psi (x).
\label{Equation3}
\end{eqnarray}%
In this representation; the Dirac spinor, $\Psi (x)$, has only two
components corresponding positive and negative energy eigenstates,
which the each one has only one spin polarization. $\overline{\sigma
}^{\mu }(x)$ are the spacetime dependent Dirac matrices and they are
written in terms of the constant Dirac matrices, $\overline{\sigma
}^{i}$, by using triads, $e_{(i)}^{\mu }(x),$ as follows;
\begin{eqnarray}
\overline{\sigma }^{\mu }(x)=e_{(i)}^{\mu }(x)\overline{\sigma
}^{i}, \label{Equation4}
\end{eqnarray}
where $\overline{\sigma }^{i}$ are the Dirac matrices in flat
spacetime and they are given by
\begin{eqnarray}
\overline{\sigma }^{i}=(\overline{\sigma }^{0},\overline{\sigma
}^{1}, \overline{\sigma }^{2})  \label{Equation5}
\end{eqnarray}
with
\begin{eqnarray}
\overline{\sigma }^{0}=\sigma ^{3}\ \ ,\overline{\sigma
}^{1}=i\sigma ^{1},\ \overline{\sigma }^{2}=i\sigma ^{2},
\label{Equation6}
\end{eqnarray}%
where $\sigma ^{1}$, $\sigma ^{2}$ \ and $\sigma ^{3}$ \ Pauli
matrices, and $\Gamma _{\mu }(x)$ are the spin affine connection by
the following definition,
\begin{eqnarray}
\Gamma_{\mu }(x)=\frac{1}{4}g_{\lambda \alpha }(e_{\nu ,\mu
}^{i}e_{i}^{\alpha }-\Gamma _{\nu \mu }^{\alpha })s^{\lambda \nu
}(x). \label{Equation7}
\end{eqnarray}%
Here, $\Gamma _{\nu \mu }^{\alpha }$ is the Christoffell symbol, and
$g_{\mu\nu}(x)$ is the metric tensor that is given in terms of the
triads as follows,
\begin{eqnarray}
g_{\mu \nu }(x)=e_{\mu }^{(i)}(x)e_{\nu }^{(j)}(x)\eta _{(i)(j)},
\label{Equation8}
\end{eqnarray}%
where $\mu $ and $\nu $ are a curved spacetime indices running from
$0$ to $2$. $i$ and $j$ are flat spacetime indices running from $0$
to $2$ and $\eta _{(i)(j)}$ is the metric of the (2+1) dimensional
Minkowski spacetime, with signature (1,-1,-1), and $s^{\lambda \nu
}(x)$ is a spin operator defined as
\begin{eqnarray}
s^{\lambda \nu }(x)=\frac{1}{2}[\overline{\sigma }^{\lambda }(x),\overline{%
\sigma }^{\nu }(x)].  \label{Equation9}
\end{eqnarray}
Using the Eq.(\ref{GUP3}), Eq.(\ref{EN}) and Eq.(\ref{MO}) in the
Dirac equation, the generalized Dirac equation becomes
\begin{eqnarray}
-i\overline{\sigma }^{0}(x)\partial _{0}\widetilde{\Psi} =\left(
i\overline{\sigma }^{i}(x)\partial _{i}-i\overline{\sigma
}^{\mu}(x)\Gamma_{\mu }-\frac{m_{0}}{\hbar}\right) \left( 1+\beta
\hbar^{2}\partial _{j}\partial ^{j}-\beta m_{0}^{2}\right)
\widetilde{\Psi}, \label{MDE1}
\end{eqnarray}
and it is rewritten as
\begin{eqnarray}
[i\overline{\sigma }^{0}(x)\partial _{0}+i\overline{\sigma
}^{i}(x)\left( 1-\beta m_{0}^{2}\right) \partial _{i}+i\beta
\hbar^{2}\overline{\sigma }^{i}(x)\partial _{i}\left(\partial
_{j}\partial ^{j}\right) -\frac{m_{0}}{\hbar}\left( 1+\beta
\hbar^{2}\partial _{j}\partial ^{j}-\beta m_{0}^{2}\right) \nonumber \\
-i\overline{\sigma }^{\mu}(x)\Gamma_{\mu}\left( 1+\beta
\hbar^{2}\partial _{j}\partial ^{j}-\beta
m_{0}^{2}\right)]\widetilde{\Psi} =0,\label{MDE2}
\end{eqnarray}
where the $\widetilde{\Psi}$ is the generalized Dirac spinor.

To calculate the tunnelling probability of a Dirac particle from the
black hole, we use the following ansatz for the wave function;
\begin{equation}
\widetilde{\Psi}(x)=\exp \left( \frac{i}{\hbar }S\left( t,r,\phi
\right) \right)\ \left(\begin{array}{c}A\left( t,r,\phi \right)
\\B\left( t,r,\phi \right) \\ \end{array}\right) \label{Equation13}
\end{equation}
where the $A\left( t,r,\phi \right) $ and $B\left( t,r,\phi \right)
$ are the functions of space-time. Inserting the
Eq.(\ref{Equation13}) in Eq.(\ref{MDE2}), we get the resulting
equations to leading order in $\hbar$ and $\beta$ as follows;
\begin{eqnarray*}
A[\frac{1}{\sqrt{F(r)}}\frac{\partial S}{\partial
t}+m_{0}\left(1-\beta m_{0}^{2}\right) +\frac{\beta m_{0}}{r^{2}}\left( \frac{\partial S}{%
\partial \phi }\right) ^{2}+\beta m_{0}F(r)\left(\frac{\partial S}{\partial
r}\right) ^{2}]\\ +B[ i\sqrt{F(r)}\left( 1-\beta
m_{0}^{2}\right)\frac{\partial S}{\partial r}+\frac{1-\beta
m_{0}^{2}}{r}\frac{\partial S}{\partial \phi }+i\beta F(r)\sqrt{F(r)}\left( \frac{\partial S}{\partial r}%
\right) ^{3}] \\+B[i\frac{\beta \sqrt{F(r)}}{r^{2}}\left( \frac{\partial S}{%
\partial r}\right) \left( \frac{\partial S}{\partial \phi }\right)
^{2}+\frac{\beta F(r)}{r}\left( \frac{\partial S}{\partial \phi }\right)
\left(\frac{\partial S}{\partial r}\right) ^{2}+\frac{\beta
}{r^{3}}\left( \frac{\partial S}{\partial \phi }\right) ^{3}] =0
\end{eqnarray*}
\begin{eqnarray}
A[-i\sqrt{F(r)}\left( 1-\beta m_{0}^{2}\right) \frac{\partial S}{%
\partial r}+\frac{1-\beta m_{0}^{2}}{r}\frac{\partial S}{\partial \phi }%
-i\beta F(r)\sqrt{F(r)}\left( \frac{\partial S}{\partial r}\right) ^{3}]\nonumber \\+A[-i%
\frac{\beta \sqrt{F(r)}}{r^{2}}\left( \frac{\partial S}{\partial
r}\right)\left( \frac{\partial S}{\partial \phi }\right) ^{2}
 +\frac{\beta F(r)}{r}\left( \frac{\partial S}{\partial
\phi }\right) \left( \frac{\partial S}{\partial r}\right) ^{2}+\frac{\beta }{r^{3}}\left( \frac{\partial S}{%
\partial \phi }\right) ^{3}] \nonumber \\ +B[ \frac{1}{\sqrt{F(r)}}\frac{%
\partial S}{\partial t}-m_{0}\left( 1-\beta m_{0}^{2}\right) -\frac{\beta
m_{0}}{r^{2}}\left( \frac{\partial S}{\partial \phi }\right)
^{2}-\beta m_{0}F(r)\left( \frac{\partial S}{\partial r}\right)
^{2}]=0\quad\;&. \label{Equation14}
\end{eqnarray}
These two equations have nontrivial solutions for the $A\left(
t,r,\phi \right)$ and $B\left(t,r,\phi \right)$ in case the
determinant of the coefficient matrix is vanished. Accordingly, when
neglecting the terms containing  higher order of the $\beta$, then
we get
\begin{eqnarray*}
\frac{1}{F(r)}\left( \frac{\partial S}{\partial t}\right)
^{2}-F(r)\left( \frac{\partial S}{\partial r}\right)
^{2}-\frac{1}{r^{2}}\left( \frac{\partial S}{\partial \phi
}\right) ^{2}-m_{0}^{2} \\+\beta [-\frac{2}{r^{4}}\left(\frac{\partial S}{\partial \phi
}\right) ^{4} -2F^{2}(r)\left( \frac{\partial S}{\partial
r}\right) ^{4} -\frac{4F(r)}{r^{2}}\left(\frac{\partial
S}{\partial r}\right) ^{2}\left( \frac{\partial S}{
\partial \phi }\right) ^{2}+2m_{0}^{4}]=0.
\end{eqnarray*}
Due to the Killing vectors $\left( \partial _{t}\right)$ and $\left(
\partial _{\phi }\right)$, we can separate the variables for
$S\left( t,r,\phi \right)$ as $S\left( t,r,\phi \right) =-Et+j\phi
+K(r)$, where, $E$ and $j$ are the energy and angular momentum of
the particle, respectively, and $K(r)=K_{0}(r)+\beta K_{1}(r)$
\cite{24k}. Then, the integral of the radial equation, $K(r)$,
becomes as follows;
\begin{eqnarray*}
K_{\pm }(r)= \pm \int \frac{\sqrt{E^{2}-F(r)\left( m_{0}^{2}+\frac{j^{2}}{%
r^{2}}\right) }}{F(r)}\left[ 1+\frac{\beta}{F(r)} \left(
\frac{2E^{2}m_{0}^{2}F(r)-E^{4}}{E^{2}-F(r)\left(m_{0}^{2}+\frac{j^{2}}{r^{2}}\right)
}\right)\right] dr
\end{eqnarray*}
and it is computed as
\begin{eqnarray*}
K_{\pm }(r_{h})=\pm i\pi \frac{l^{2}E}{972B^{3}}\left[
216B^{2}+\beta \left( 324m_{0}^{2}B^{2}+16E^{2}l^{2}-27j^{2}\right)
\right]
\end{eqnarray*}
Thus, from the Eq.(\ref{proba1}) and Eq.(\ref{proba2}), the
tunneling probability of the Dirac particle is given by
\begin{eqnarray*}
\Gamma =\frac{P_{out}}{P_{in}}=\exp \left\{ -\frac{\pi l^{2}E}{243\hbar B^{3}}%
\left[ 216B^{2}+\beta \left( 324m_{0}^{2}B^{2}+16E^{2}l^{2}-27j^{2}\right) %
\right] \right\}.
\end{eqnarray*}
Furthermore, from the Eq.(\ref{proba2}), the modified Hawking
temperature becomes as follows
\begin{eqnarray}
T_{H}^{^{\prime }}&=&\frac{9 \hbar B}{8\pi l^{2}}\left[ 1+\beta \frac{%
324m_{0}^{2}B^{2}+16E^{2}l^{2}-27j^{2}}{216B^{2}}\right] ^{-1} \nonumber  \\
&=&T_{H}\left[ 1-\beta \frac{%
324m_{0}^{2}B^{2}+16E^{2}l^{2}-27j^{2}}{216B^{2}}\right],
\label{DTH}
\end{eqnarray}
where the $T_{H}=\frac{9\hbar B}{8\pi l^{2}}$ is the standard
Hawking temperature of the Martinez-Zanelli black hole. As in the
case of the scalar particle tunnelling, the corrected Hawking
temperature of the tunnelling Dirac particle is related to not only
the mass parameter of the Martinez-Zanelli black hole, but also
depends the angular momentum, energy and mass of the emitted Dirac
particle, and it is lower than the standard Hawking temperature.

\section{Concluding remarks}\label{conc}

In this paper we have studied the issue of the quantum gravity
effect on the Hawking radiation of the 2+1 dimensional
Martinez-Zanelli black hole by using the particle tunnelling method.
To take into account the quantum gravity effects, we modified the
Dirac and Kelin-Gordon equations by the generalized fundamental
commutation relations to discuss the tunneling radiation of fermions
and scalar particles, respectively. The results showed that the
corrected Hawking temperature is not only determined by the mass
parameter of the Martinez-Zanelli black hole, but also it is
affected by the quantum properties (i.e., the angular momentum,
energy and mass) of the emitted fermions and scalar particles. The
other important results are given as follows:
\begin{itemize}
\item According to Eq.(\ref{KGTH}), the corrected Hawking
temperature of the tunnelling scalar particle is lower than the
standard temperature.
\item In Eq.(\ref{DTH}), when
$324m_{0}^{2}B^{2}+16E^{2}l^{2}>27j^{2}$, the corrected Hawking
temperature of the tunnelling fermions is lower than the standard
temperature. However, when $324m_{0}^{2}B^{2}+16E^{2}l^{2}<27j^{2}$,
the corrected temperature is higher than the standard temperature.
If $324m_{0}^{2}B^{2}+16E^{2}l^{2}=27j^{2}$, then the contribution
of the GUP effect is canceled, and radiation temperature of the
tunnelling fermions reduce to the standard temperature.
\item By comparing the Eq.(\ref{DTH} with Eq.(\ref{KGTH}), we can say that
the radiation temperature of the tunnelling fermions higher than the
scalar particles temperature, even if their masses, energies, and
angular momentums are same.
\end{itemize}

Finally, thanks to the GUP effect, we can determine whether the
radiated particle from a black hole is the scalar particle or the
Dirac particle.

\section*{Acknowledgments}

This work was supported by the Scientific Research Projects Unit of
Akdeniz University.


\end{document}